\documentclass[a4paper,11pt]{article}
\usepackage{pos}
\usepackage{bbm}
\usepackage{slashed}
\usepackage{subcaption}
\newcommand{\MSbar}{{\overline{\rm MS}}}

\title{Perturbative study of renormalization and mixing for asymmetric staple-shaped Wilson-line operators on the lattice }
\ShortTitle{Renormalization of asymmetric staple-shaped Wilson-line operators}

\author*[a]{Gregoris Spanoudes}
\author[b]{Martha Constantinou}
\author[a]{Haralambos Panagopoulos}

\affiliation[a]{Department of Physics, University of Cyprus, P.O. Box 20537, 1678 Nicosia, Cyprus}

\affiliation[b]{Department of Physics, Temple University, Philadelphia, Pennsylvania 19122 - 1801, USA}

\emailAdd{gspano01@ucy.ac.cy}
\emailAdd{marthac@temple.edu}
\emailAdd{haris@ucy.ac.cy}

\abstract{We present one-loop perturbative results of the renormalization functions for a complete set of nonlocal quark bilinear operators containing an asymmetric staple-shaped Wilson line, using a family of improved lattice actions. This study is relevant for the nonperturbative investigations regarding the renormalization of the unpolarized, helicity and transversity transverse-momentum dependent parton distribution functions (TMDPDFs) in lattice QCD. We employ a number of different versions of regularization-independent (RI$'$) renormalization prescriptions which address the power and logarithmic divergences of such nonlocal operators, the pinch-pole singularities at infinite Wilson-line lengths, as well as the mixing among operators of different Dirac structures, as dictated by discrete symmetries. All cancelations of divergences and admixtures are confirmed by our results at one-loop level. We compare all the different prescriptions and we provide the conversion matrices at one-loop order which relate the matrix elements of the staple operators in RI$'$ to the reference scheme $\MSbar$.
\begin{center}
\includegraphics[scale=0.45]{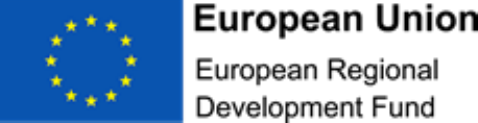}
\includegraphics[scale=0.45]{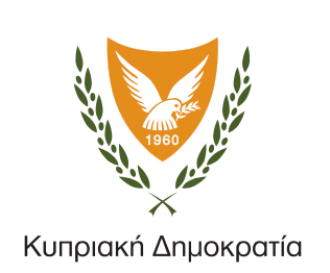}
\includegraphics[scale=0.45]{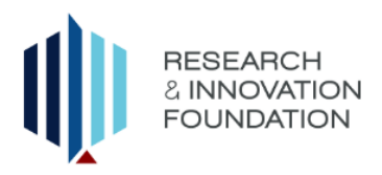}
\includegraphics[scale=0.45]{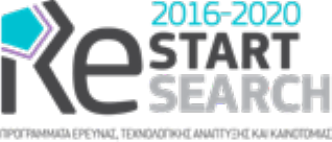}
\end{center}} 

\FullConference{The 40th International Symposium on Lattice Field Theory (Lattice 2023)\\
July 31st - August 4th, 2023\\
Fermi National Accelerator Laboratory\\}

\begin{document}
\maketitle

\section{Introduction}
One of the directions of research in lattice QCD, which shows a rapid progress in recent years, is the study of transverse momentum-dependent parton distribution functions (TMDPDFs). These functions encode the dependence of hadrons on the transverse -- to the direction of movement -- momentum of their constituents (valence and sea quarks, gluons), providing insights in the investigation of the three-dimensional hadron structure. While collinear PDFs, which probe the one-dimensional structure of hadrons, have been extensively studied in both experiments and theory, the study of TMDPDFs has been relatively limited. This status is expected to change in the coming years with new data coming from the planned electron ion colliders of USA and China. Theoretical calculations of TMDPDFs using lattice QCD can significantly complement the experimental efforts by providing input into the analysis and interpretation of the experimental data.

The calculation of PDFs, TMDPDFs, etc., from first principles has long been a challenge in Hadron Physics due to their nonperturbative and light-cone nature. The latter does not allow for their direct nonperturbative computation on a Euclidean lattice. In the last decade, a pioneering method for extracting these quantities on the lattice was suggested by X. Ji. This method is based on the calculation of Euclidean equal-time correlation functions (called quasi-PDFs, quasi-TMDPDFs, etc.), which are accessible by lattice simulations, and their connection to the physical light-cone distribution functions using the framework of Large Momentum Effective Theory (LaMET). A crucial intermediate step in this method is the renormalization of the quasi-distribution functions, which is necessary for making contact to physically measurable quantities. 

The quasi-distribution functions are hadron matrix elements of gauge-invariant nonlocal operators, which contain path-ordered Wilson lines of different shapes. In this work, we focus on the renormalization of nonlocal quark bilinear operators with an asymmetric staple-shaped Wilson line, relevant for the TMDPDFs. There are a number of challenges to address in order to renormalize such operators properly, including power-law divergences, logarithmic divergences which stem from the singular points of the Wilson line (cusps, end points), mixing between operators with different Dirac structures, and pinch-pole singularities in the infinite limit of lateral sizes of the staple.

In what follows, we employ one-loop perturbation theory in both dimensional (DR) and lattice (LR) regularizations and we calculate  the renormalization functions for a complete set of asymmetric staple-shaped operators. We consider a number of different regularization-independent (RI$'$) prescriptions based on the allowed operator mixing as dictated by symmetries. Moreover, we provide the one-loop conversion matrices which match the RI$'$-renormalized operators (for each variant) to the $\MSbar$ scheme, employed in phenomenology. A long write-up of our work, together with an extended list of references, can be found in a forthcoming paper \cite{Spanoudes:2023}.

\section{Calculation Setup}
\label{Sec2}
\begin{figure}[htb]
    \centering
    \includegraphics[width=0.8\textwidth]{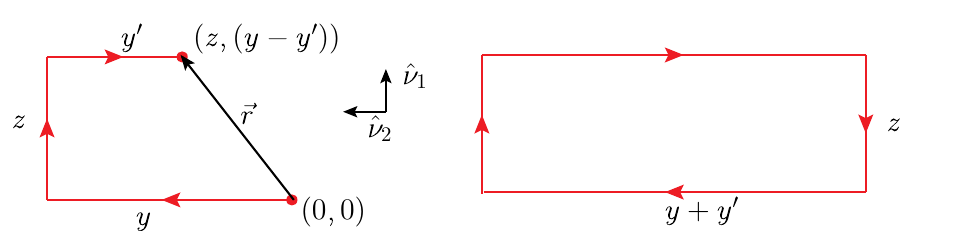}
    \caption{The shapes of $\mathcal{W}_{\rm staple} (z,y,y')$ (left) and $\mathcal{W}_{\rm loop} (z,y + y')$ (right).}
    \label{fig:staple_line}
  \end{figure}

A set of 16 independent nonlocal operators of different Dirac matrices are considered in this study, which are defined in Euclidean space as follows:
\begin{equation}
    \mathcal{O}_\Gamma (z,y,y') \equiv \Bar{\psi}(0) \ \Gamma \ \mathcal{W}_{\rm staple} (z, y, y') \ \psi (z \hat{\nu}_1 + (y-y') \hat{\nu}_2),
    \label{O_Gamma}
\end{equation}
where $\mathcal{W}_{\rm staple} (z,y,y')$ denotes the path-ordered staple-shaped Wilson line as given schematically in Fig.~\ref{fig:staple_line}, and  $\Gamma = \mathbbm{1}, \gamma_5, \gamma_{\nu_i}, \gamma_5 \gamma_{\nu_i}, \sigma_{\nu_i \nu_j} \equiv [\gamma_{\nu_i}, \gamma_{\nu_j}]/2, \ (i=1,2,3,4)$. $\hat{\nu}_1, \hat{\nu_2}$ are orthogonal directions which form the plane in which the staple lies (see Fig.~\ref{fig:staple_line}) and $\hat{\nu}_3, \hat{\nu}_4$ are orthogonal directions perpendicular to this plane.

One of the main features of the nonlocal Wilson-line operators is that finite mixing among operators of different Dirac structures can occur on the lattice. It is first realized in Ref. \cite{Constantinou:2017sej}, in which a one-loop perturbative calculation of straight Wilson-line operators on the lattice has been performed by our group. This calculation shows that certain pairs of straight Wilson-line operators must be renormalized as mixing doublets in order to match the lattice bare Green's functions to the $\MSbar$ scheme. In the same spirit, a follow-up one-loop lattice calculation for the case of symmetric staple-shaped Wilson-line operators shows mixing in pairs with a different pattern compared to the case of straight Wilson line. However, symmetry arguments can show a wider mixing in both symmetric and asymmetric staple operators; this is examined in the current work. 

We investigate the properties of asymmetric staple operators $\mathcal{O}_\Gamma$ under the symmetries of parity $\mathcal{P}$, time reversal $\mathcal{T}$, charge conjugation $\mathcal{C}$ and chiral transformations. A generalization of both parity and time reversal in each direction is obtained in Euclidean space. The symmetry transformations of $\mathcal{O}_\Gamma$ are (translational symmetry has been also applied): 
\begin{eqnarray}
\mathcal{O}_{\Gamma} (z,y,y') &\xrightarrow[]{\mathcal{P}_{\mu}}& \mathcal{O}_{\gamma_{\mu} \Gamma \gamma_{\mu}} ((-1)^{\delta_{\mu \nu_1} + 1} \ z, (-1)^{\delta_{\mu \nu_2} + 1} \ y, (-1)^{\delta_{\mu \nu_2} + 1} \ y'), \\
\mathcal{O}_{\Gamma} (z,y,y') &\xrightarrow[]{\mathcal{T}_{\mu}}& \mathcal{O}_{\gamma_5 \gamma_{\mu} \Gamma \gamma_{\mu} \gamma_5} ((-1)^{\delta_{\mu \nu_1}} \ z, (-1)^{\delta_{\mu \nu_2}} \ y, (-1)^{\delta_{\mu \nu_2}} \ y'), \\
\mathcal{O}_{\Gamma} (z,y,y') &\xrightarrow[]{\mathcal{C}}& \mathcal{O}_{(C \Gamma C^{-1})^{T}} (-z,y',y),
\end{eqnarray}
where $C$ is the charge conjugation matrix satisfying $C \gamma_\mu C^{-1} = - \gamma^T_\mu$. Under chiral transformations, $\mathcal{O}_\Gamma$ is invariant only for $\Gamma = \gamma_{\mu}, \gamma_5 \gamma_{\mu}$. By taking appropriate combinations of staple operators, which are odd/even under $\mathcal{C}, \mathcal{P}, \mathcal{T}$ \cite{Alexandrou:2023ucc}, the following conclusions for the operator mixing are obtained: When the fermion action breaks the chiral symmetry, mixing among the quadruplets of asymmetric staple operators $(\mathcal{O}_{\Gamma}, \mathcal{O}_{ \Gamma \gamma_{\nu_1} \gamma_{\nu_2}}, \mathcal{O}_{\Gamma \gamma_{\nu_1}}, \mathcal{O}_{\Gamma \gamma_{\nu_2}})$ is allowed. In the case of chiral fermions, the quadruplets are reduced to pairs $(\mathcal{O}_{\Gamma}, \mathcal{O}_{\Gamma \gamma_{\nu_1} \gamma_{\nu_2}})$. In the specific case of symmetric staple operators ($y' = y$), the mixing pattern is simplified to $(\mathcal{O}_\Gamma, \mathcal{O}_{[\Gamma, \gamma_{\nu_1} \gamma_{\nu_2}]/2}, \mathcal{O}_{[\Gamma, \gamma_{\nu_1}]/2}, \mathcal{O}_{[\Gamma, \gamma_{\nu_2}]/2})$ and $(\mathcal{O}_\Gamma, \mathcal{O}_{[\Gamma, \gamma_{\nu_1} \gamma_{\nu_2}]/2})$ for non-chiral and chiral fermions, respectively. Depending on the commutation relations of each Dirac matrix $\Gamma$ with $\gamma_{\nu_1}$, $\gamma_{\nu_2}$, and $\gamma_{\nu_1} \gamma_{\nu_2}$, the mixing multiplets become triplets (doublets) or singlets for the case of non-chiral (chiral) fermions. 

In our study, we employ four different variants of the regularization independent (RI$'$) scheme for renormalizing the asymmetric staple-shaped operators. All prescriptions consider $4 \times 4$ renormalization matrices $Z_{\Gamma \Gamma'}^{X,R}$ (where $X (R)$ denotes the regularization (renormalization) scheme) respecting the wider mixing dictated by symmetries. The mixing sets of operators $\mathcal{O}_\Gamma$ are written explicitly in terms of $\Gamma$ matrices: $S_1 \equiv  \{ \mathbbm{1}, \sigma_{\nu_1 \nu_2}, \gamma_{\nu_1},\gamma_{\nu_2} \}$, $S_2 \equiv \{ \gamma_5, \sigma_{\nu_4 \nu_3}, \gamma_5 \gamma_{\nu_1}, \gamma_5 \gamma_{\nu_2} \}$, $S_3 \equiv \{ \gamma_{\nu_3}, \gamma_5 \gamma_{\nu_4}, \sigma_{\nu_3 \nu_1},\sigma_{\nu_3 \nu_2} \}$, $S_4 \equiv \{ \gamma_{\nu_4}, \gamma_5 \gamma_{\nu_3}, \sigma_{\nu_4 \nu_1}, \sigma_{\nu_4 \nu_2} \}$. 

The first two schemes, referred to as RI$'_i$ ($i=1,2$), are defined by the condition\footnote{From now on, sums over repeated $\Gamma$ matrices are understood.}:
\begin{equation}
  \frac{1}{4 N_c} {(Z^{X,{\rm RI}'}_\psi)}^{-1}  Z^{X,{\rm RI}'_i}_{\Gamma \Gamma'} {\rm Tr}[\Lambda_{\Gamma'} (q,z,y,y') P^{[i]}_{\Gamma''}] \Big|_{q = \bar{q}} = \delta_{\Gamma \Gamma''}, \quad (\Gamma, \Gamma'' \in S_j, \ i=1,2, \ j=1-4),
  \end{equation}  
where $\bar{q}$ is the RI$'$ renormalization 4-vector scale, $\Lambda_\Gamma (q,z,y,y') = {\langle \psi (q)| \mathcal{O}_\Gamma (z,y,y') | \bar{\psi} (q) \rangle}_{\rm amp.}$ is the amputated Green's function of $\mathcal{O}_\Gamma$ with external fermion fields $\psi(q)$, and $Z^{X,R}_\psi$ is the renormalization factor of $\psi(q)$. We use the conventions: $\mathcal{O}^R_\Gamma (z,y,y') = Z^{X,R}_{\Gamma \Gamma'} \mathcal{O}_{\Gamma'} (z,y,y')$, and $\psi^R (q) = {(Z^{X,R}_\psi)}^{1/2} \psi (q)$. The projector $P^{[i]}_{\Gamma}$ is defined for each scheme as:
\begin{eqnarray}
P_{\Gamma}^{[1]} &=& \quad \ e^{-i {\sf q} \cdot {\sf r}} \ \Gamma^\dagger, \\
  P_{\Gamma}^{[2]} &=& 
  \begin{cases}
  e^{-i {\sf q} \cdot {\sf r}} \left(\mathbbm{1} - \frac{\slashed{q}_T \ \slashed{q}_L}{q_T^2} \right) \Gamma^\dagger, & \qquad \Gamma \in S_1, S_2 \\
  e^{-i {\sf q} \cdot {\sf r}} \left(\mathbbm{1} - \frac{(\slashed{q}_T - \slashed{q}_{\nu_3}) (\slashed{q}_L + \slashed{q}_{\nu_3})}{q_T^2 - q_{\nu_3}^2} \right)\Gamma^\dagger, & \qquad \Gamma \in \{\gamma_{\nu_3}, \gamma_5 \gamma_{\nu_3}, \sigma_{\nu_3 \nu_1}, \sigma_{\nu_3 \nu_2}\} \\
  e^{-i {\sf q} \cdot {\sf r}} \left(\mathbbm{1} - \frac{(\slashed{q}_T - \slashed{q}_{\nu_4}) (\slashed{q}_L + \slashed{q}_{\nu_4})}{q_T^2 - q_{\nu_4}^2} \right)\Gamma^\dagger, & \qquad \Gamma \in \{\gamma_{\nu_4}, \gamma_5 \gamma_{\nu_4}, \sigma_{\nu_4 \nu_1}, \sigma_{\nu_4 \nu_2}\}
  \end{cases},
\end{eqnarray}
 where $\vec{\sf r} \equiv z \ \hat{\nu}_1 + (y-y') \ \hat{\nu}_2$, $\vec{q}_L \equiv q_{\nu_1} \hat{\nu}_1 + q_{\nu_2} \hat{\nu}_2$ and $\vec{q}_T \equiv \vec{q} - \vec{q}_L = q_{\nu_3} \hat{\nu}_3 + q_{\nu_4} \hat{\nu}_4$. Compared to the first projector, the second one can further remove finite contributions of some Dirac structures, allowed by residual rotational symmetry, from the elements of the renormalization matrices.

 Even though the above two schemes can address all renornalization issues in one-loop perturbation theory, recent nonperturbative investigations \cite{Zhang:2022xuw} at multiple lattice spacings show remaining linear divergences, as well as significant nonperturbative effects at large distances. An alternative prescription, in order to overcome these issues, considers modified operators $\bar{\mathcal{O}}_\Gamma \equiv \mathcal{O}_{\Gamma}/{\langle \mathcal{W}_{\rm loop} (z, y+y')\rangle}^{1/2}$, where $\langle \mathcal{W}_{\rm loop} (z,y+y') \rangle$ is the vaccuum expectation value of a rectangular Wilson loop given in Fig.~\ref{fig:staple_line}. Then the condition takes the following form, resulting to two alternative schemes, referred to as RI$'_1$-bar and RI$'_2$-bar, ($\bar{\mathcal{O}}^R_\Gamma = \bar{Z}^{X,R}_{\Gamma \Gamma'} \bar{\mathcal{O}}_{\Gamma'}$):
 \begin{equation}
  \frac{1}{4 N_c} {(Z^{X,{\rm RI}'}_\psi)}^{-1} \bar{Z}^{X,{\rm RI}'_i}_{\Gamma \Gamma'} \frac{{\rm Tr}[\Lambda_{\Gamma'} (q,z,y,y') P^{[i]}_{\Gamma''}]}{\sqrt{\langle \mathcal{W}_{\rm loop} (z, y+y')\rangle}} \Big|_{q = \bar{q}} = \delta_{\Gamma \Gamma''}, \quad (\Gamma, \Gamma'' \in S_j, \ i=1,2, \ j=1-4).
  \end{equation} 
The square root of $\langle \mathcal{W}_{\rm loop} \rangle$ can cancel the linear, cusp and pinch-pole divergences of the operators $\mathcal{O}_\Gamma$. Cancellations of the pinch-pole divergences allows one to take the limit $y \rightarrow \infty$ in the renormalized Green's functions of $\bar{\mathcal{O}}_\Gamma$ in any scheme (including $\MSbar$). Then $\bar{Z}^{X,R}_{\Gamma \Gamma'}$ addresses the remaining endpoint divergences, as well as the mixing. Since the endpoint divergences do not depend on the dimensions of the staple, a nonperturbative determination of $\bar{Z}^{X,R}_{\Gamma \Gamma'}$ is expected to exhibit a much milder dependence on the staple lengths $z,y, y'$. In this way it becomes more acceptable to renormalize the modified operators $\bar{\mathcal{O}}_{\Gamma}$ defined at large values of the lengths $z,y, y'$, using renormalization functions $\bar{Z}^{X,RI'}_{\Gamma \Gamma'}$ defined at smaller values of $z,y, y'$ within the perturbative region. Hence, nonperturbative effects at large distances are suppressed. Lattice discretization effects stemming from the use of small values of the staple parameters can be reduced by subtractions of artifacts calculated in one-loop perturbation theory.

\section{Calculation in dimensional and lattice regularizations}

\begin{figure}[htb]
  \centering
   \includegraphics[width=0.95\textwidth]{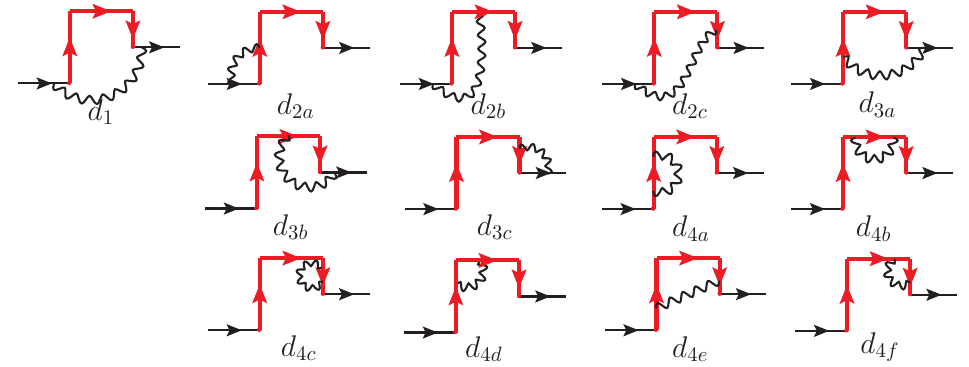}
    \caption{One-loop diagrams contributing to $\Lambda_\Gamma$. The straight (wavy) lines represent fermions (gluons).}
    \label{fig:one-loop_diagrams}
\end{figure}

The Feynman diagrams that enter our one-loop calculations are shown in Fig.~\ref{fig:one-loop_diagrams}. These diagrams will appear in both DR and LR, since all vertices are present in both regularizations. Due to the non-local nature of the staple-shaped operators, multiple scales (staple lengths) appear in the Green's functions, which make the computation more complex even at one-loop level. The corresponding calculation on the lattice is even more demanding since the procedure for isolating divergences from the Feynman integrals, as well as the procedure for taking the continuum limit $a \rightarrow 0$ (where $a$ is the lattice spacing) are more complicated (see, e.g., \cite{Constantinou:2017sej}).

The perturbative calculation in DR ($D \equiv 4-2\varepsilon$ dimensions) demonstrates that pole terms ($\mathcal{O} (1/\varepsilon)$) in the Green's functions $\Lambda_\Gamma$ are multiples of the tree-level values and therefore there is no mixing in applying $\MSbar$ to DR.  We identified that these $\mathcal{O} (1/\varepsilon)$ terms come from the sum of cusp ($d_{4d}$, $d_{4f}$), endpoint ($d_{2a}$, $d_{3c}$) and contact singularities ($d_{4a}$, $d_{4b}$, $d_{4c}$).The multiplicative renormalization factors for the operators $\mathcal{O}_\Gamma$ and $\bar{\mathcal{O}}_\Gamma$ in the $\MSbar$ scheme are given by:
\begin{equation}
    Z^{{\rm DR},{\overline{\rm MS}}}_{\Gamma} = 1 - \frac{(g^{\overline{\rm MS}})^2}{16 \pi^2} C_F \frac{7}{\varepsilon} + \mathcal{O} \left((g^{\overline{\rm MS}})^4\right), \quad \bar{Z}^{{\rm DR},{\overline{\rm MS}}}_{\Gamma} = 1 - \frac{(g^{\overline{\rm MS}})^2}{16 \pi^2} C_F \frac{3}{\varepsilon} + \mathcal{O} \left((g^{\overline{\rm MS}})^4\right).
    \label{ZMSbar}
\end{equation}
They are independent of the Dirac matrix $\Gamma$, and the lengths of the staple. The finite terms in $\Lambda_\Gamma$ are complex functions with a nontrivial dependence on the staple lengths; they are expressed in terms of integrals over Feynman parameters, which involve Bessel functions. We have also identified the  pinch-pole singularity, which comes from a term $[y/z \tan^{-1} (y/z) + y'/z \tan^{-1} (y'/z)]$ of diagram $d_{4e}$, when $y \rightarrow \infty$ is taken for fixed values of $y-y'$. This term is cancelled (in the infinite $y$-limit) by a term $(y+y')/z \tan^{-1} ((y+y')/z)$ appearing in $\langle \mathcal{W}_{\rm loop} \rangle$, when the modified operator $\bar{\mathcal{O}}_\Gamma$ is employed. The full expressions of the bare Green's functions can be found in \cite{Spanoudes:2023}. 

\begin{figure}[htb]
    \centering  
    \begin{subfigure}{0.49\textwidth}
    \includegraphics[width=\textwidth]{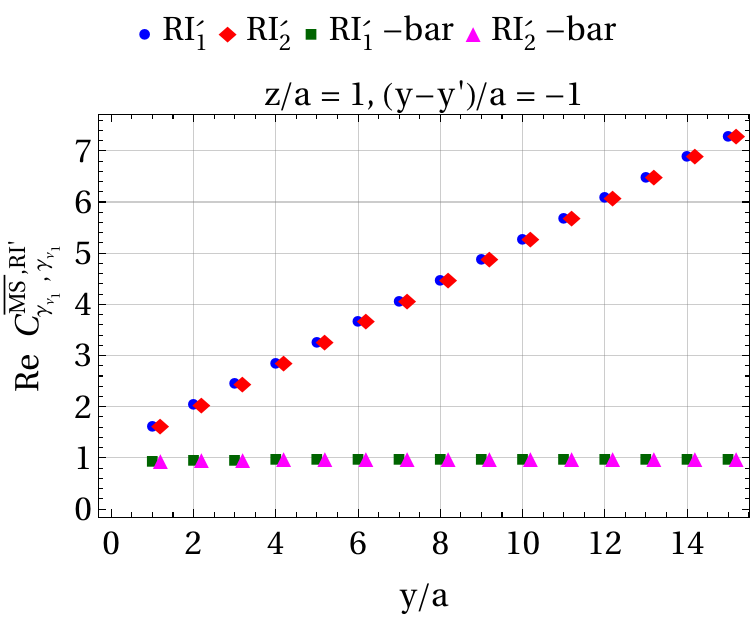}
    \end{subfigure}
    \begin{subfigure}{0.5\textwidth}
    \includegraphics[width=\textwidth]{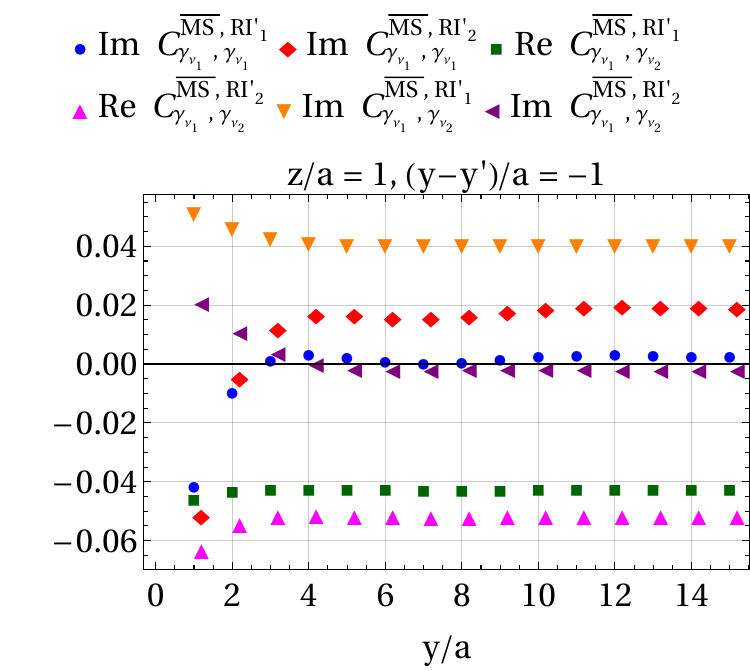}
    \end{subfigure}
    \caption{Conversion matrix elements for $\mathcal{O}_{\gamma_{\nu_1}}$ as functions of $y/a$ [$\bar{\mu} = 2$ GeV ($\MSbar$ scale), $\beta = 1$ (Landau gauge), $a \bar{q} = (\frac{2 \pi}{L} n_1, \frac{2 \pi}{L} n_2, \frac{2 \pi}{L} n_3, \frac{2 \pi}{T} (n_4 + \frac{1}{2}))$, $n_1=n_2=n_3=4$, $n_4=5$, $L=32$, $T=64$ and $a=0.09$ fm].}
    \label{fig:Cfactors}
\end{figure}

An important outcome of this perturbative study is the calculation of the one-loop regularization-independent conversion matrices which relate results from the RI$'$-type schemes (described in Sec.~\ref{Sec2}) to the reference scheme of $\MSbar$. The conversion matrices take a $4 \times 4$ block-diagonal form (see \cite{Spanoudes:2023}). In Fig.~\ref{fig:Cfactors}, we illustrate some representative conversion matrix elements relevant for the operator $\mathcal{O}_{\gamma_{\nu_1}}$. We compare all four RI$'$-type schemes employed in our work. Note that the imaginary parts, as well as the nondiagonal elements, are identical between RI$'_i$ and RI$'_i$-bar schemes at one loop. As expected, the real part of the diagonal elements which appear in the left plot has an almost linear (flat) dependence on $y/a$ in the RI$'_i$ (RI$'_i$-bar) schemes. This is related to the presence/absence of the pinch-pole divergence. Differences between the schemes with index 1 and 2 are visible only on the elements of the right plot. However, the contribution from these elements is much milder compared to the real diagonal parts of the left plot.

The perturbative calculation in LR (Wilson/clover fermions, Symanzik-improved gluons) allows for extracting one-loop results for a general Wilson-line lattice operator with $n$ cusps when considering the difference between its bare lattice and $\MSbar$-renormalized Green’s functions $\delta \Lambda_\Gamma$:
\begin{eqnarray}
    \delta\Lambda_\Gamma &=& - \frac{g^2\,C_F}{16\,\pi^2}\, e^{i\,{\sf q} \cdot {\sf r}} \Big\{ 2 \ \Gamma \left[ \boldsymbol{\alpha_1} + 16 \pi^2 P_2 \, \beta + (1 - \beta) \log (a^2 \bar{\mu}^2)\right] + \frac{1}{2} (\Gamma \hat{\slashed{\nu}}_i + \hat{\slashed{\nu}}_{\hspace{-0.4mm}f} \Gamma) (\boldsymbol{\alpha_2} + \boldsymbol{\alpha_3} c_{SW}) \nonumber \\
    &&  + \Gamma \Big[ (n + 1) \boldsymbol{\alpha_4} + n \boldsymbol{\alpha_5} - 16 \pi^2 P_2 \, \beta + \left( 2 (n + 1) + \beta \right) \log (a^2 \bar{\mu}^2) + \boldsymbol{\alpha_6} \frac{\ell}{a} \Big] \Big\} + \mathcal{O} (g^4),
    \label{general_WL}
\end{eqnarray}
where $\ell$ is the length of the Wilson line, $c_{SW}$ is the clover coefficient, $\beta$ is the gauge-fixing parameter [$\beta = 1 \ (0)$: Landau (Feynman) gauge], $\bar{\mu}$ is the $\MSbar$ renormalization scale and $\hat{\nu}_{_i} \ (\hat{\nu}_{_{\hspace{-0.5mm}f}})$ is the direction of the Wilson line in the initial (final) end point. $P_2 = 0.0240$ and $\alpha_i$ are given in Table~\ref{tab:stapleLR} for different gluon actions (Wilson, Tree-Level Symanzik, Iwasaki).
\begin{table}[thb]
  \centering
  \begin{tabular}{|l|l|l|l|l|l|l|}
  \hline
\ \textbf{Gluon action} & \ \quad \ $\boldsymbol{\alpha_1}$ & \ \quad \ $\boldsymbol{\alpha_2}$ & \ \quad \ $\boldsymbol{\alpha_3}$ & \ \quad \ $\boldsymbol{\alpha_4}$ & \ \quad \ $\boldsymbol{\alpha_5}$ & \ \quad \ $\boldsymbol{\alpha_6}$\\
\hline
\hline
\ Wilson & \ -4.4641 \ & \ \ \ 14.4499 \ & \ \ -8.2847 \ & \ -4.5258 \ & \ \ \ \ \ \ 0 \ & \ \ 19.9549 \ \\
\ TL Symanzik \ & \ -4.3413 \ & \ \ \ 12.7559 \ & \ \ -7.6736 \ & \ -3.9303 \ & \ -0.8099 \ & \ \ 17.2937 \ \\ 
\ Iwasaki & \ -4.1637 \ & \ \ \ \phantom{0}9.9365 \ & \ \ -6.5276 \ & \ -1.9053 \ & \ -2.1011 \ & \ \ 12.9781 \ \\
\hline
  \end{tabular}
  \caption{Numerical values of the coefficients $\alpha_1 - \alpha_6$ appearing in Eq. \eqref{general_WL}. }
  \label{tab:stapleLR}
\end{table}
The generalization on the shape of the Wilson line comes from the fact that any parts of the Wilson line which do not include singular points will give finite contributions, which vanish in $\delta \Lambda_\Gamma$ when taking the continuum limit. We identify all types of singularities, including linear divergences ($\ell/a$) which depend on the total length of the Wilson line, cusp and endpoint logarithmic divergences ($\log (a^2 \bar{\mu}^2)$, the coefficients of which depend on the number of cusps. We also identify mixing contributions $(\Gamma \hat{\slashed{\nu}}_i + \hat{\slashed{\nu}}_{\hspace{-0.4mm}f} \Gamma)/2$, which depend on the direction of the staple line entering the end points and come from the chirality-breaking parts of the action. However, as concluded by symmetries, additional mixing structures are expected beyond one loop. 

In order to investigate RI$'$-bar schemes on the lattice, we have also calculated the one-loop bare Green’s function of $\langle \mathcal{W}_{\rm loop}\rangle$ and we confirm the cancellation of the linear and cusp divergences between the staple operator ($n=2$) and the Wilson loop. The lattice renormalization matrices of $\mathcal{O}_\Gamma$ and $\bar{\mathcal{O}}_\Gamma$ in the $\MSbar$ scheme can be easily extracted by our calculations:
\begin{eqnarray}
    Z^{{\rm LR},\MSbar}_{\Gamma \Gamma'} &=& \delta_{\Gamma \Gamma'} \Big[1 - \frac{(g^\MSbar)^2 C_F}{16 \pi^2} \Big(e (2) + e_1^{\psi} + e_2^{\psi} c_{\rm SW} + e_3^{\psi} c_{\rm SW}^2 - \alpha_6 \frac{|z|+|y|+|y'|}{a} - 7 \ln (a^2 \bar{\mu}^2) \Big)\Big]\nonumber \\
    && + \delta_{\Gamma', [\Gamma, \gamma_{\nu_2}]/2} \ {\rm sgn}(y) \ \frac{(g^\MSbar)^2 C_F}{16 \pi^2} \left(\alpha_2 + \alpha_3 c_{\rm SW}\right) + \mathcal{O} ((g^{\overline{\rm MS}})^4), \\
     \bar{Z}^{{\rm LR},\MSbar}_{\Gamma \Gamma'} &=& \delta_{\Gamma \Gamma'} \Big[1 - \frac{(g^\MSbar)^2 C_F}{16 \pi^2} \Big( e(0) + e_1^{\psi} + e_2^{\psi} c_{\rm SW} + e_3^{\psi} c_{\rm SW}^2 - 3 \ln (a^2 \bar{\mu}^2) \Big)\Big] \nonumber \\
&& + \delta_{\Gamma', [\Gamma, \gamma_{\nu_2}]/2} \ {\rm sgn}(y) \ \frac{(g^\MSbar)^2 C_F}{16 \pi^2} \left(\alpha_2 + \alpha_3 c_{\rm SW}\right) + \mathcal{O} ((g^{\overline{\rm MS}})^4).
\end{eqnarray}
where $e(n) \equiv 1 - 2 \alpha_1 - (n+1) \alpha_4 - n \alpha_5$ and $e^{\psi}_i$ comes from $Z_\psi^{{\rm LR}, \MSbar}$.

\section{Summary - Future plans}
In our work, we have identified the mixing pattern among asymmetric staple-shaped Wilson-line operators using symmetry arguments for both chiral and non-chiral fermions. We confirm in one-loop perturbation theory that appropriate RI$'$-type schemes address all type of divergences and operator mixing in both continuum and lattice regularizations. One-loop conversion matrices to the $\MSbar$ scheme are calculated for staple operators of different Dirac matrices. A future plan is to extend our work to two loops, which will be useful for improving the current nonperturbative investigations. \\

\hspace{-0.7cm}{\bf \large Acknowledgments} \\
This work is funded by the European Regional Development Fund and the Republic of Cyprus through the Research and Innovation Foundation (EXCELLENCE/0421/0025). M.C. acknowledges financial support by the U.S. Department of Energy, Office of Nuclear Physics, Early Career Award under Grant No. DE-SC0020405.


\bibliographystyle{JHEP}
\bibliography{refs2.bib}

\end{document}